\documentclass[useAMs,a4paper]{mn2e}
\usepackage{savesym}
\usepackage{graphicx}
\expandafter\let\csname equation*\endcsname\relax
  \expandafter\let\csname endequation*\endcsname\relax 
\usepackage{subfig}
\usepackage{amsmath}
\usepackage{amssymb}
\usepackage{verbatim}
\usepackage[yyyymmdd,hhmmss]{datetime}
\usepackage{array}
\usepackage{times}
\usepackage[total={17.8cm,24.0cm},centering]{geometry} 
\usepackage{color}

\newcommand{\beq}{\begin{equation}}
\newcommand{\eeq}{\end{equation}}

\newcommand\T {{\widetilde {T}}}

\newcommand\ind{{  \eta  }}

\title[The  high energy disc luminosity distribution] {The high energy probability distribution of accretion disc luminosity fluctuations }
\author [Andrew Mummery, Steven Balbus]{Andrew Mummery\thanks{E-mail:
andrew.mummery@physics.ox.ac.uk}, Steven Balbus
\\
Oxford Astrophysics, Denys Wilkinson Building, Keble Road, Oxford, OX1 3RH, United Kingdom}
\begin{document}

\date{}

\pagerange{\pageref{firstpage}--\pageref{lastpage}} \pubyear{2022}

\maketitle

\label{firstpage}

%%%%----------------------------------------------%%%%----------------------------------------------%%%%----------------------------------------------%%%%----------------------------------------------%%%%
\begin{abstract} 
The probability density function of accretion disc luminosity fluctuations at high observed energies (i.e., energies larger than the peak temperature scale of the disc) is derived, under the assumption that the temperature fluctuations are log-normally distributed.   Thin disc theory is used throughout.     While log-normal temperature fluctuations would imply that the disc's {\it bolometric} luminosity is also log-normal, the observed Wien-like luminosity behaves very differently.   For example, in contrast to a log-normal distribution,  the standard deviation of the derived distribution is not linearly proportional to its mean. This means that these systems do not follow a linear rms-flux relationship. Instead they exhibit very high intrinsic variance, and undergo what amounts to a phase transition, in which the mode of the distribution (in the statistical sense) ceases to exist, even for physically reasonable values of the underlying temperature variance. The moments of this distribution are derived using asymptotic expansion techniques.  A result that is important for interpreting  observations is that the theory predicts that the fractional variability of these disc systems should increase as the observed frequency is increased. The derived distribution  will be of practical utility in quantitatively understanding the variability of disc systems observed at energies above their peak temperature scale, including X-ray observations of tidal disruption events.   
\end{abstract}

\noindent
%Complied at \today\ \currenttime\ .
%%%%----------------------------------------------%%%%----------------------------------------------%%%%----------------------------------------------%%%%----------------------------------------------%%%%

\section{Introduction}
Accretion discs are fundamentally turbulent systems, a result of the magnetorotational instability (Balbus \& Hawley 1991). As such, the observed emission from all accretion disc sources generally exhibits temporal variability.  The power spectral densities of the luminosity emergent from accretion discs reveal fluctuations which are found to be log-normally distributed.   This was first shown by Uttley {\it et al.} (2005), using the observed linear relationship between the root-mean square (rms) variation on short timescales and the mean flux varying over longer timescales (Uttley \& McHardy 2001). Uttley {\it et al}. (2005) showed that if this linear rms-flux relationship extends to all timescales (as suggested by observational data), then the corresponding light curve must be log-normally distributed.  

Despite the vast range of both length and time scales involved, the variability of accretion disc systems display remarkable similarity in their properties across a broad population of sources. In any individual source, variability as rapid as the local evolutionary timescale of the innermost disc edge is observed. Variability is also observed at all timescales longer than this, up to many orders of magnitude longer than any physical process which is involved in the direct production of the disc luminosity.  
Log-normal variability properties are also observed in systems which span a vast array of length scales: from compact discs in Galactic X-ray binaries (e.g., Gleissner et al. 2004), to the very large discs in active galactic nuclei (e.g., Vaughan et al. 2011). This behaviour has been confirmed with observations at various different  frequencies in black hole accretion disc sources, including both AGN in X-ray (Gaskell 2004; Vaughan et al. 2011), AGN in optical (Lyutyi \& Oknyanskii 1987), and X-ray binaries at both X-ray (Gleissner et al. 2004) and optical (Gandhi 2009) frequencies.  We also note that non black hole disc sources, e.g., cataclysmic variables (Scaringi et al. 2012), and young stellar objects (Scaringi et al. 2015) also show the same variability structure.

A model that has been put forward to explain these observed properties is the so-called ``theory of propagating fluctuations'', first described by Lyubarskii (1997).  The basic assumption here is that the standard $\alpha$-prescription (Shakura \& Sunyaev 1973) for the effective disc turbulence should be modified with small stochastic perturbations (driven on the viscous timescale) added to the value of $\alpha$ at each disc radius. These $\alpha$ perturbations then result in surface density $\Sigma$ fluctuations (or equivalently mass accretion rate $\dot M$ fluctuations) at each radius, which evolve through the disc. This model successfully reproduces many of the properties of variable accretion systems (e.g., Turner \& Reynolds 2021).  The log-normality of the resulting disc bolometric light curve is a robust prediction of this model. While MHD simulations find that the $\alpha$ parameter is itself log-normally distributed (Hogg \& Reynolds 2016), it was recently shown that even if the fluctuations in $\alpha$ were assumed to be normally distributed, the resulting $L_{\rm bol}$ fluctuations was still found to be log-normal (Turner \& Reynolds 2021). 

There is ample evidence therefore, both observational (e.g., Uttley {\it et al}. 2005) and theoretical (e.g., Turner \& Reynolds 2021), that the bolometric luminosity of an accretion disc is log-normally  distributed.  However, one never observes the true bolometric luminosity; instead, one observes the luminosity across some fixed instrumental bandpass.   If this bandpass samples from the bulk of the disc spectrum, then it is indeed likely that the observed luminosity will also follow a log-normal distribution. However, this is far from always the case, and it is in fact not true in a rather common physical limit: when observations are taken at energies higher than the peak energy scale of the disc itself.   

To understand this point more fully, consider the expression for the observed thermal ``soft state'' emission from an accretion disc observed across a band pass from $\nu_l$ to $\nu_h$:
\beq\label{fullmodel}
L_{\rm band} =  4 \pi \int_{\nu_l}^{\nu_h} \iint_{\cal S} {f_\gamma^3 f_{\rm col}^{-4} B_\nu (\nu/f_\gamma, f_{\rm col} T)} ~\text{d}b_1\text{d} b_2 \, {\rm d}\nu.
\eeq
Here ${\cal S}$ is the surface of the disc, $B_\nu$ is the Planck function,  and $b_1$ and $b_2$ are cartesian {image plane} photon impact parameters. The factor $f_\gamma$ is the photons energy-shift factor, defined as the ratio of the observed photon frequency $\nu$ to the emitted photon frequency $\nu_{\rm emit}$, $f_\gamma \equiv \nu/\nu_{\rm emit}$.  $T$ is the temperature of the disc, potentially a function of both disc radius and time $T(r,t)$. Finally, $f_{\rm col}$ is the `color-correction' factor, which  models disc {opacity} effects. 
(For a more detailed derivation  see e.g. Mummery \& Balbus 2021a.) 

While this expression is generally complex and unwieldy, in the limit when the inner disc is cool compared to the observing frequency,  the formal disc spectrum peaks below the lower band pass of the instrument, $kT  \ll h \nu_l$, and the integral of interest becomes analytically tractable. In Mummery \& Balbus (2021a) it was shown that right side of equation (\ref{fullmodel}) can be evaluated by performing a Laplace expansion about the hottest region in the disc, resulting in the following expression 
\begin{equation}\label{MB}
L_{\rm band}  = \frac{16 \pi^2 \xi_1 h\nu_l^4}{c^2f_{\rm col}^4} R_p^2 \left(\frac{k \T_p}{h \nu_l} \right)^\ind \exp\left(- \frac{h\nu_l}{k \T_p} \right) .
\end{equation} 
Here we have defined $\T_p \equiv f_{\rm col} f_\gamma T_p$, where $T_p$ is the hottest temperature in the accretion disc. The radius $R_p$ corresponds to the radial image plane coordinate of the peak temperature, while the constant $\ind$ depends on the inclination angle of the disc and the disc's inner boundary condition, and is limited to the range $3/2 \leq \ind \leq  5/2$. The constant $\xi_1 \simeq 2.19$.  

The luminosity emergent from an accretion disc, in the Wien-like limit $h \nu_l \gg k \T_p$, depends exponentially on the disc temperature.   Thus, if the disc temperature is itself log-normally distributed (as it must be if the bolometric luminosity is log-normally distributed), then $L_{\rm band}$ {\it will not be log-normally distributed}.  Under these conditions, how is the high photon energy luminosity  distributed?   We answer this question in this paper.  It turns out that the distribution function of $L_{\rm band}$ can be written in a closed form in the  limit of direct interest. This distribution has a number of important mathematical properties which are very distinctive and important for interpreting observations, and which have no analogue in the log-normal distribution.    

{Zdziarski (2005) also discussed some of the variability  properties of an accretion disc's Wien-tail emission, but allowed only the flux amplitude  to fluctuate, while keeping the shape (and therefore the temperature) of the spectrum constant.    This approach misses the remarkable behaviour stemming from the mathematical properties of the exponentiations of the turbulent temperature fluctuations.  These are essential to include in order to describe what lies behind observational variability, in addition to being physically well motivated. }

A particularly relevant example of a Wien-tail disc system arises in the study of tidal disruption events, which have typical peak temperatures of order $kT_p \sim 50-150$ eV (e.g. Brown {\it et al}. 2017), and are observed at X-ray energies (typically, $h \nu_l = 300$ eV).   The variability properties of tidal disruption event X-ray light curves should therefore be described by the probability distribution derived in this paper, and not the log-normal distribution.   These two distributions are very different.    {Note, however, that the analysis of this paper focusses on disc systems whose emission is locally thermal, and as such will be less appropriate for comparison to classical AGN spectra, whose X-ray emission is often dominated by a hot corona. }

We begin with a derivation of the probability density function of the Wien-like disc luminosity.  We will then analyse its properties in some detail. 

\section {Variability of the high energy disc luminosity}\label{lnf}

\subsection{The luminosity probability distribution function }
We have seen that the high energy luminosity distribution can be described by a (dimensionless) function of the form:
\beq\label{dimX}
Y = X^\ind \, \exp(-1/X) \equiv g(X), 
\eeq
where $Y = L_{\rm band}/L_0$,  $X = f_{\rm col} T_p/T_l$, and $L_0$ and $T_l$ are disc temperature independent constants defined as
 \begin{align}\label{constdef}
L_0 &\equiv {16\pi^2 \xi_1 \over c^2 f^4_{\rm col} } h \nu_l^4 R_p^2,  \\
T_l &\equiv {h \nu_l \over k} .
 \end{align}
For convenience, when producing numerical plots we shall assume that $R_p$ takes the value of the innermost stable circular orbit (ISCO) of the Schwarzschild spacetime, a value which is likely to be reasonably accurate $R_p = {6 G M_{\rm BH} /c^2}$, where $M_{\rm BH}$ is the black hole mass. The dimensionless disc temperature  $X$ is taken to be distributed log-normally,  with probability density $p_X(x)$. The probability density function of the luminosity variable $Y$ is defined 
\beq
p_Y(y) = {\partial \over \partial y} \Bigg[{\rm Pr} (Y \leq y)\Bigg]
\eeq
where ${\rm Pr}(Y\leq y)$ is the cumulative probability function of the variable $Y$. This can be related to the known distribution $p_X(x)$ via 
\beq
{\rm Pr} (Y \leq y) = {\rm Pr}(g(X) \leq y ) = {\rm Pr}(X \leq g^{-1}(y) ),
\eeq
where the final equality follows from the fact that $Y = g(X)$ is a monotonically increasing function with a well-defined 
inverse, $X=g^{-1}(Y)$.   Therefore,
\begin{multline}\label{formalPydef}
p_Y(y) = {\partial \over \partial y} \Bigg[{\rm Pr} (X \leq g^{-1}(y))\Bigg] \\ = p_X(g^{-1}[y]) \,\,{\partial \over \partial y} \Big( g^{-1}[y] \Big) .
\end{multline}
Thus, to compute the distribution of the luminosity, we must invert equation (\ref{dimX}) for the inverse function\footnote{Here and throughout this paper, the $-1$ exponent in $g^{-1}$  denotes an inverse function.  In all other appearances, it denotes a reciprocal.} $g^{-1}(Y)$.  Equation (\ref{dimX}) may be rearranged in the form 
\beq
{1 \over \ind Y^{1/\ind}} =  {1 \over \ind X} \exp\left( {1 \over \ind X}\right) ,
\eeq
which can then be inverted for $X(Y)$:
\beq
X =  g^{-1}(Y)= 1 \Big/ \left( \ind W\left[\ind^{-1} Y^{-1/\ind} \right] \right).
\eeq
In this expression $W$ is the Lambert W function  (Corless et al. 1996), 
which is defined as the solution $w$ of the following equation 
\beq\label{lambdef}
w e^w = z \rightarrow w \equiv W(z). 
\eeq
The derivative of the Lambert W function has the following useful property 
\beq\label{derivlamb}
{{\rm d} W(z) \over {\rm d} z} = {1 \over z + \exp\left(W\right)} = {\exp(-W) \over 1 + W} ,
\eeq
which follows directly from its definition (equation [\ref{lambdef}]). The derivative we require for computing the luminosity probability density function is
\beq
{{\rm d} X \over {\rm d} Y} =  \left({1 \over \ind}\right)^2 {z \over Y} \left({1 \over W\left(z\right)}\right)^{2} {{\rm d} W(z) \over {\rm d} z},% \left({1 \over \ind}\right)^3 Y^{-(1+\ind)/\ind} \left(W\left[\ind^{-1} Y^{-1/\ind}\right]\right)^{-2} {{\rm d} W(z) \over {\rm d} z},
\eeq
where we have defined $z \equiv \ind^{-1} Y^{-1/\ind}$. Using equations (\ref{lambdef}) and (\ref{derivlamb}) to simplify, we are left with 
\beq
{{\rm d} X \over {\rm d} Y} = { \ind^{-2} \, \, Y^{-1} \over W(z)\left(1 + W(z)\right) } .%{ \ind^{-2} \, \, Y^{-1} \over W\left[\ind^{-1} Y^{-1/\ind}\right]\left(1 + W\left[\ind^{-1} Y^{-1/\ind}\right]\right) } .
\eeq

We may now construct the luminosity distribution function.    Start with the probability density function of temperature fluctuations, which follows a log-normal profile, and has the following form:
\beq\label{pln}
p_X (x; \mu_N, \sigma_N) = {1\over \sqrt{2\pi}\sigma_N x }\exp (-[\ln(x)-\mu_N]^2/2\sigma_N^2),
\eeq
where this function is valid only for $x > 0$ (positive temperatures), and $\mu_N$ and $\sigma_N^2$ are the mean and variance of the underlying normal distribution (these will be related to the mean $\mu_T$ and variance $\sigma_T^2$ of the temperature distribution shortly). The Lambert W function satisfies 
\beq\label{loglamb}
\ln\left(W(z)\right) = \ln(z) - W(z), 
\eeq
a result which follows from its definition (equation [\ref{lambdef}]). Substituting $X(Y)$ into $p_X(x)$ gives the first term in our required expression:
\begin{multline}
p_X(y; \ind, \mu_N, \sigma_N) = {\ind W(z) \over \sqrt{2\pi} \sigma_N} \times \\ 
\exp\left[- \left(W(z) + \ind^{-1}\ln(y) - \mu_N\right)^2 \Big/ 2\sigma_N^2\right] .
 % {\ind W\left[\ind^{-1} y^{-1/\eta}\right] \over \sqrt{2\pi} \sigma_N} \times \\ 
%\exp\left(- \left(W\left[\ind^{-1}y^{-1/\ind}\right] + \ind^{-1}\ln(y) - \mu_N\right)^2 \Big/ 2\sigma_N^2\right) .
\end{multline}
Combining with ${\rm d}X/{\rm d}Y$ from above, we are left with 
\begin{multline}\label{xray_dist}
p_Y(y; \ind, \mu_N, \sigma_N) = {\ind^{-1} y^{-1} \over \sqrt{2\pi} \sigma_N \left(1 + W(z) \right) } \times \\
\exp\left[- \left(W(z) + \ind^{-1}\ln(y) - \mu_N\right)^2 \Big/ 2\sigma_N^2\right] ,
 %{\ind^{-1} y^{-1} \over \sqrt{2\pi} \sigma_N \left(1 + W\left[\ind^{-1}y^{-1/\ind}\right] \right) } \times \\
%\exp\left(- \left(W\left[\ind^{-1}y^{-1/\ind}\right] + \ind^{-1}\ln(y) - \mu_N\right)^2 \Big/ 2\sigma_N^2\right) ,
\end{multline}
which is the exact probability density function of the normalised luminosity variable $Y$, the key result of this paper. Recall that the true luminosity variable $L_{\rm band}$ is related to the variable $Y$ by a simple scaling $L_{\rm band} = L_0 Y$, and that $L_0$ depends on the system parameters only through the black hole mass $M_{\rm BH}$ (which sets the radial scale $R_p$ in equation [\ref{constdef}]). The high energy luminosity distribution is therefore a four-parameter distribution which may be found by combining equation (\ref{xray_dist}) with the definition of $L_0$ (equation [\ref{constdef}]): 
\beq
p_{L}(l; M_{\rm BH}, \ind, \mu_T, \sigma_T) = L_0^{-1} \, p_Y(y = l/L_0; \ind, \mu_N, \sigma_N) ,
\eeq 
where the mean ($\mu_N$) and variance ($\sigma_N^2$) of the underlying normal distribution which appear in equation (\ref{xray_dist}), are related to the  mean ($\mu_T$) and variance ($\sigma_T^2$) of the temperature distribution by 
\begin{align}
\mu_N &= \ln\left({\mu_T \over T_l} {\mu_T \over \sqrt{\mu_T^2 + \sigma_T^2}}\right) ,\label{normmu} \\
\sigma_N^2 &= \ln\left(1 + {\sigma_T^2 \over \mu_T^2}\right) . \label{varmu}
\end{align}

\begin{figure}
\includegraphics[width=\linewidth]{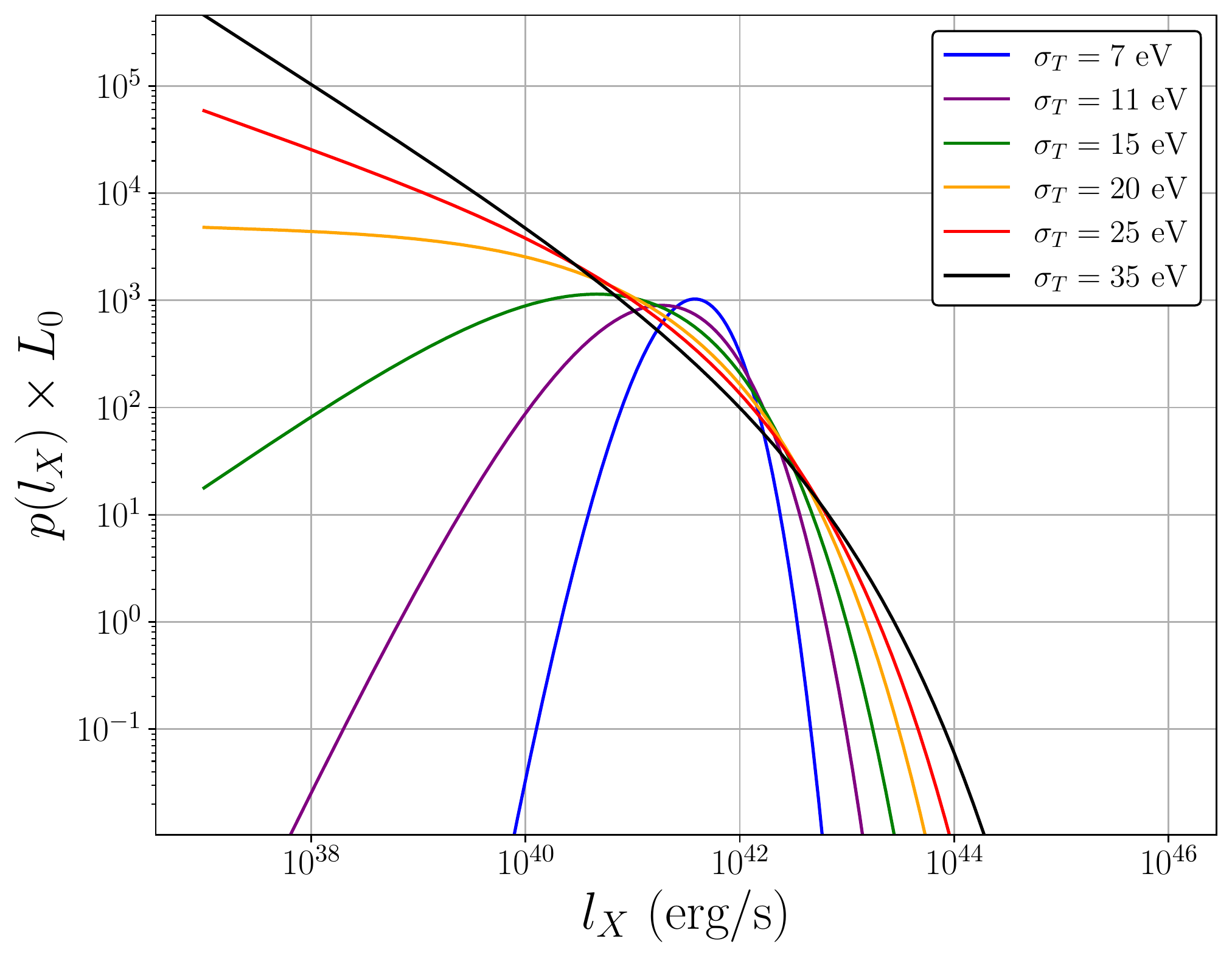}
\caption{ The X-ray luminosity probability density function, multiplied by the constant X-ray luminosity scale $L_0$ of eq. (\ref{constdef}), plotted against the X-ray luminosity. This figure was created for mean temperature  $\mu_T = 70$eV. The probability density function is an extremely sensitive function of the temperature variance $\sigma_T$, undergoing a transition from a distribution with well-defined mode to a mode-less distribution at $\sigma_T/\mu_T \sim 0.28$.   }
\label{PDF_1}
\end{figure}

\begin{figure}
\includegraphics[width=\linewidth]{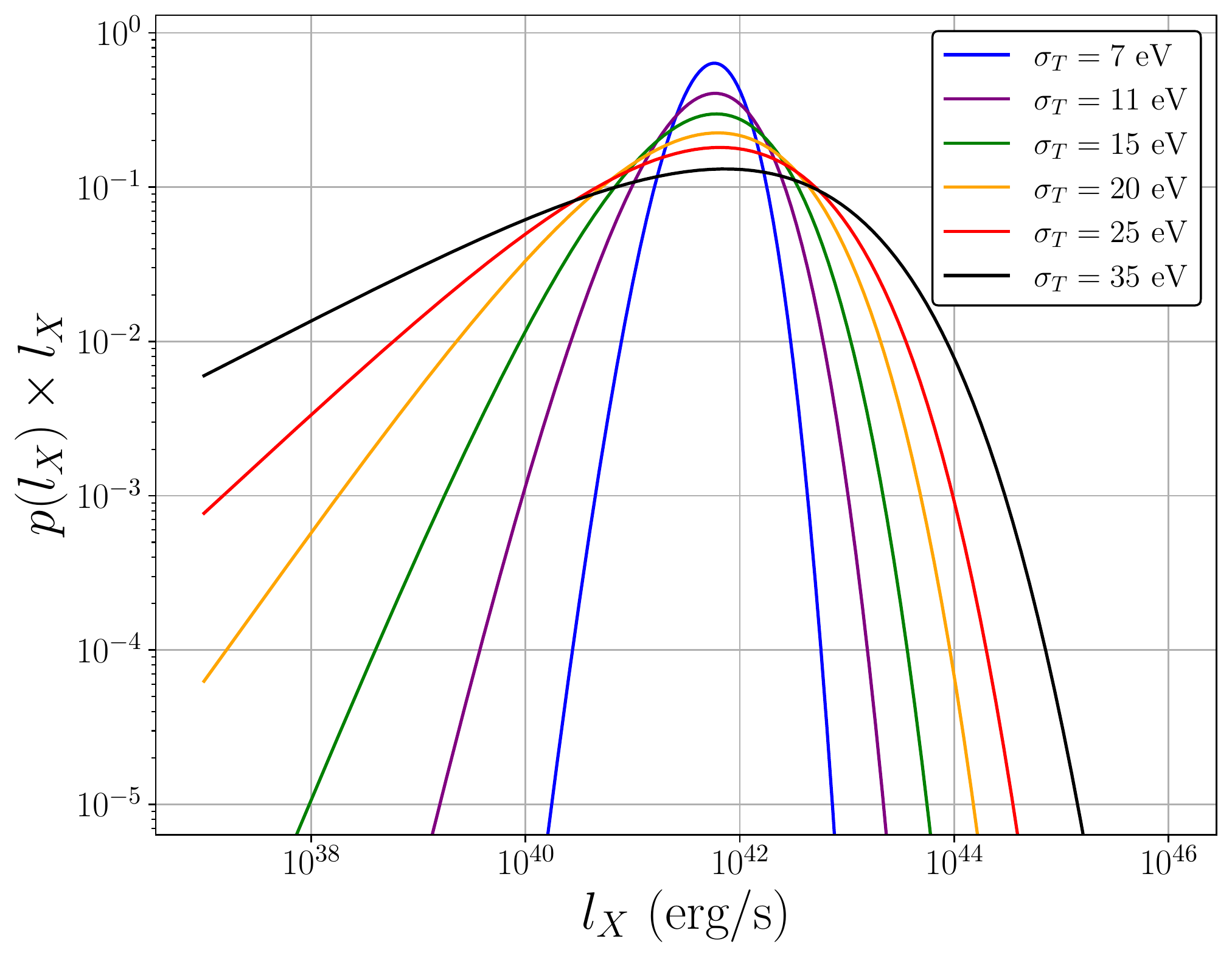}
\caption{  The X-ray luminosity probability density function, multiplied by the  X-ray luminosity  $l_X$, plotted against the X-ray luminosity. This figure was created for mean temperature  $\mu_T = 70$eV. The probability density function is an extremely sensitive function of the temperature variance, $\sigma_T$.  }
\label{PDF_2}
\end{figure}

\begin{figure}
\includegraphics[width=\linewidth]{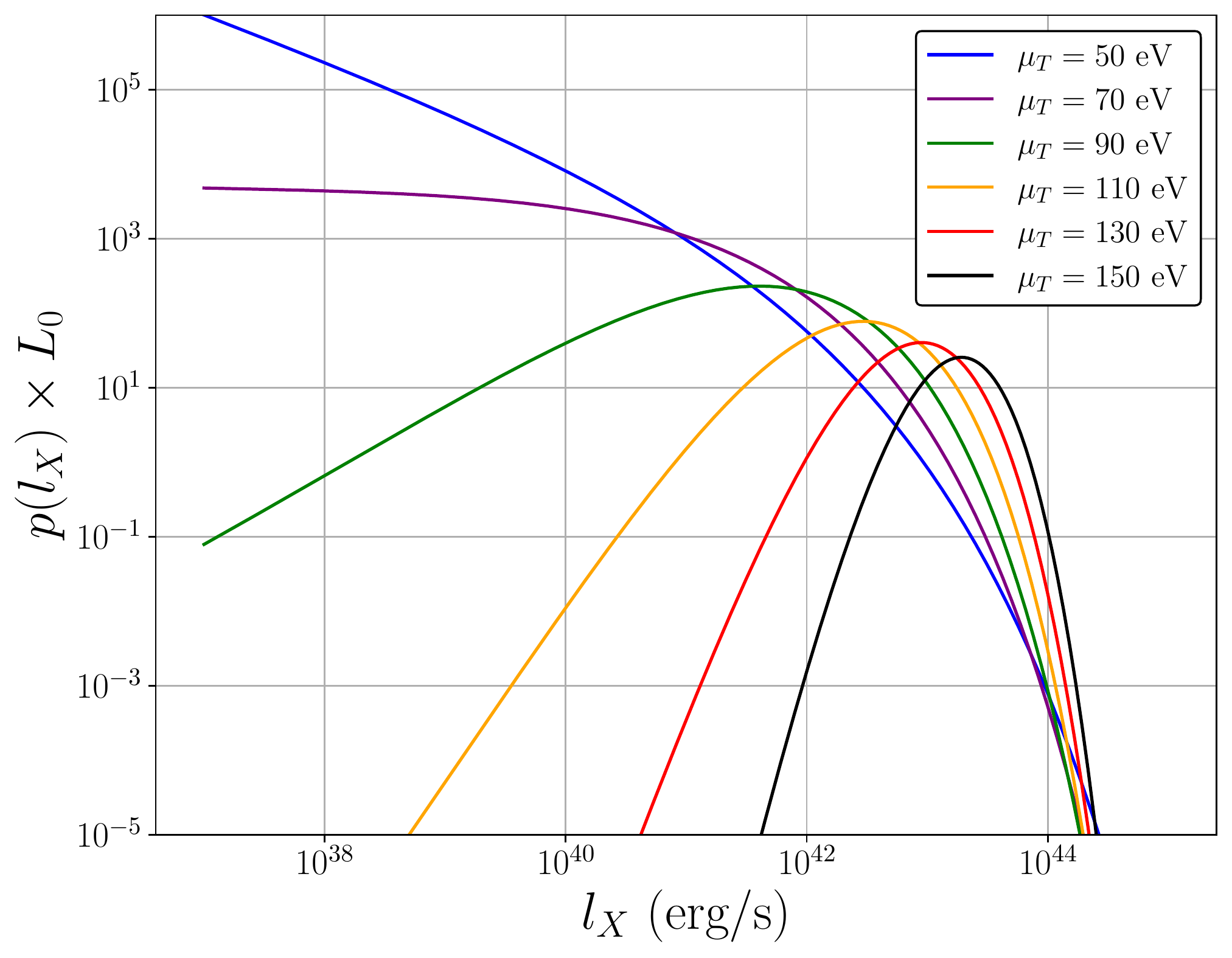}
\caption{ The X-ray luminosity probability density function, multiplied by the constant X-ray luminosity scale $L_0$, plotted against the X-ray luminosity. This figure was created with a temperature variance parameter of $\sigma_T = 20$eV. The probability density function is a sensitive function of the temperature mean, $\mu_T$.   }
\label{PDF_3}
\end{figure}

\begin{figure}
\includegraphics[width=\linewidth]{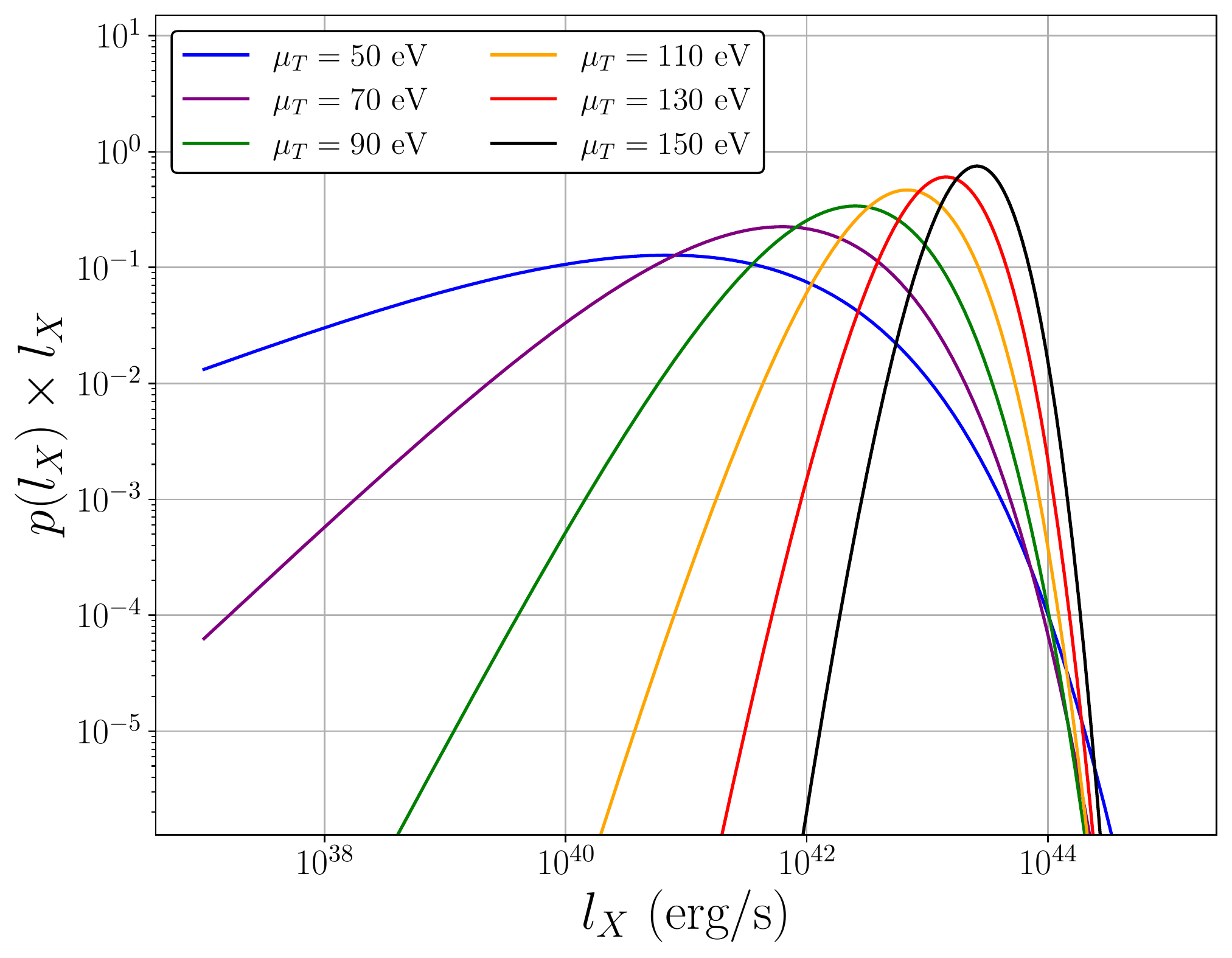}
\caption{  The X-ray luminosity probability density function, multiplied by the  X-ray luminosity  $l_X$, plotted against the X-ray luminosity. This figure was created  with a temperature variance parameter of $\sigma_T = 20$eV. The probability density function is an  sensitive function of the temperature mean, $\mu_T$.  }
\label{PDF_4}
\end{figure}

Some of the properties of this  probability density function are illustrated in Figs.\ [\ref{PDF_1}-\ref{PDF_4}].  In these plots, we assume typical values of parameters relevant for tidal disruption event X-ray light curves, which are an important observational application of this analysis. We therefore take $T_l = 300$ eV so that the luminosity variable corresponds to a typical X-ray luminosity of a tidal disruption event, which we denote $L_X$. 

In Fig. [\ref{PDF_1}] the probability density function, scaled by the constant $L_0$, is plotted against the X-ray luminosity $l_X$ for a number of different temperature variances $\sigma_T$.  Each plot was made assuming $\mu_T = 70$ eV, a representative temperature scale for a TDE system (e.g., Brown {\it et al}. 2017), and black hole mass $M_{\rm BH} = 10^6 M_\odot$.  It is clear from Fig.  [\ref{PDF_1}] that the properties of the X-ray luminosity distribution are extremely sensitive to the variance of the underlying temperature distribution.  Indeed, for temperature distributions with high intrinsic scatter, in this case $\sigma_T/\mu_T \gtrsim 0.2$, the probability density function of the X-ray luminosity rises strongly at low values and the distribution has no formal mode. In this model, high amplitude dimming events of TDE X-ray light curves are an intrinsic feature and should be a common observational occurrence.   

In Fig.\  [\ref{PDF_2}], we plot the combination $p_{L_X} l_X$ against the X-ray luminosity for the same parameter values as Fig. [\ref{PDF_1}].  Clearly $p_{L_X}$ rises less rapidly than $1/l_X$ for low values of $l_X$ and high temperature variance.  The combination $p_{L_X} l_X$ goes to zero at both high and low $l_X$, but the variance of these light curves is extreme. As an explicit example, the change in the combination $p_{L_X} l_X$ is only a factor of $\sim 10$ over {\it six orders of magnitude in $l_X$} for the highest temperature variance curve in Fig. [\ref{PDF_2}]. 

In reality, it is chiefly the ratio $\sigma_T/\mu_T$ which dominates the observed high energy scatter.   This can be seen in Figs. [\ref{PDF_3}] and [\ref{PDF_4}].    Both figures assume a constant temperature scatter $\sigma_T = 20$ eV, a black hole mass $M = 10^6 M_\odot$, and a range of mean temperatures $\mu_T$.  While Fig. [\ref{PDF_4}] shows that the mean of the high energy distribution increases with mean temperature in the expected fashion (i.e., larger mean temperatures lead to larger mean luminosities), both Figs. [\ref{PDF_3}] and [\ref{PDF_4}] demonstrate that the mean of the temperature distribution also has an important role to play in determining the variance of the luminosity distribution.  Low temperature discs (which includes all transient disc systems at late times, once their discs have cooled) show a much higher luminosity variance.

\subsection{The moments of the  luminosity distribution }
A probability distribution is generally characterised by a series of summary statistics, e.g., the mean, variance, and skewness.  To compute these summary statistics, various moments of the distribution are required.  For the distribution (\ref{xray_dist}) these integrals can in fact be solved explicitly by using  asymptotic expansion techniques, as we now show.  

The moments of the high energy distribution are given by the integral 
\begin{multline}
I_\alpha \equiv \left\langle L^\alpha \right\rangle = \int l^\alpha \, p_{L}(l) \, {\rm d}l = L_0^{\alpha}  \int y^\alpha \, p_Y(y) \, {\rm d}y  \\ = L_0^{\alpha} \int y^\alpha  \,p_X(g^{-1}[y]) \,\,{\partial \over \partial y} \Big( g^{-1}[y] \Big) \, {\rm d }y ,
\end{multline}
where $\alpha = 0, 1, 2, \dots$ labels the moment, and we have used the formal definition of equation (\ref{formalPydef}) to derive the final equality. Following the substitution $x = g^{-1}(y)$ we have 
\beq
I_\alpha = L_0^{\alpha}\int \left[g(x)\right]^\alpha p_X (x) \, {\rm d}x .
\eeq
Substituting in the log-normal distribution from equation (\ref{pln}),
\beq
I_\alpha = {L_0^{\alpha}\over \sqrt{2\pi}\sigma_N} \int\limits_0^\infty x^{\alpha\ind}  \exp \left(-{[\ln(x)-\mu_N]^2\over2\sigma_N^2} -{\alpha\over x}\right) \, {{\rm d}x \over x}, 
\eeq
and changing variable to $z = \ln(x)$, we obtain
\beq
I_\alpha = {L_0^{\alpha} \over \sqrt{2\pi}\sigma_N} \int\limits_{-\infty}^{\infty} \exp\left(-{(z - \mu_N)^2 \over 2\sigma_N^2} + \alpha \ind z - \alpha e^{-z}\right) {\rm d} z .
\eeq
After completing the square this becomes 
\begin{multline}
I_\alpha = {L_0^{\alpha} \over \sqrt{2\pi}\sigma_N} \exp\left(\alpha \ind \mu_N + {1\over 2} \alpha^2 \ind^2 \sigma_N^2\right) \\ \int\limits_{-\infty}^{\infty} \exp\left(-{(z - \mu_N - \alpha \ind \sigma_N^2)^2 \over 2\sigma_N^2}  - \alpha e^{-z}\right) {\rm d} z . 
%\equiv   {L_0^{\alpha} \over \sqrt{2\pi} \sigma_N} \exp\left(\alpha \ind \mu_N + {1\over 2} \alpha^2 \ind^2 \sigma_N^2\right) f_\alpha(\mu_N, \sigma_N, \ind) .
\end{multline}
Computing the moments of the distribution amounts to computing the properties of the dimensionless functions  $f_\alpha(\mu_N, \sigma_N, \ind)$
\beq\label{fdef}
f_\alpha  = \int\limits_{-\infty}^{\infty} \exp\left(-{(z - \mu_N - \alpha \ind \sigma_N^2)^2 \over 2\sigma_N^2}  - \alpha e^{-z}\right) {\rm d} z .
\eeq
Other than for the case $\alpha = 0$ (in which case $f_0 = \sqrt{2\pi}\sigma_N$),  $f_\alpha$ does not have a closed form solution in terms of elementary functions.   However, highly accurate asymptotic approximations of this integral may be found, as we now demonstrate.

\begin{figure}
\includegraphics[width=\linewidth]{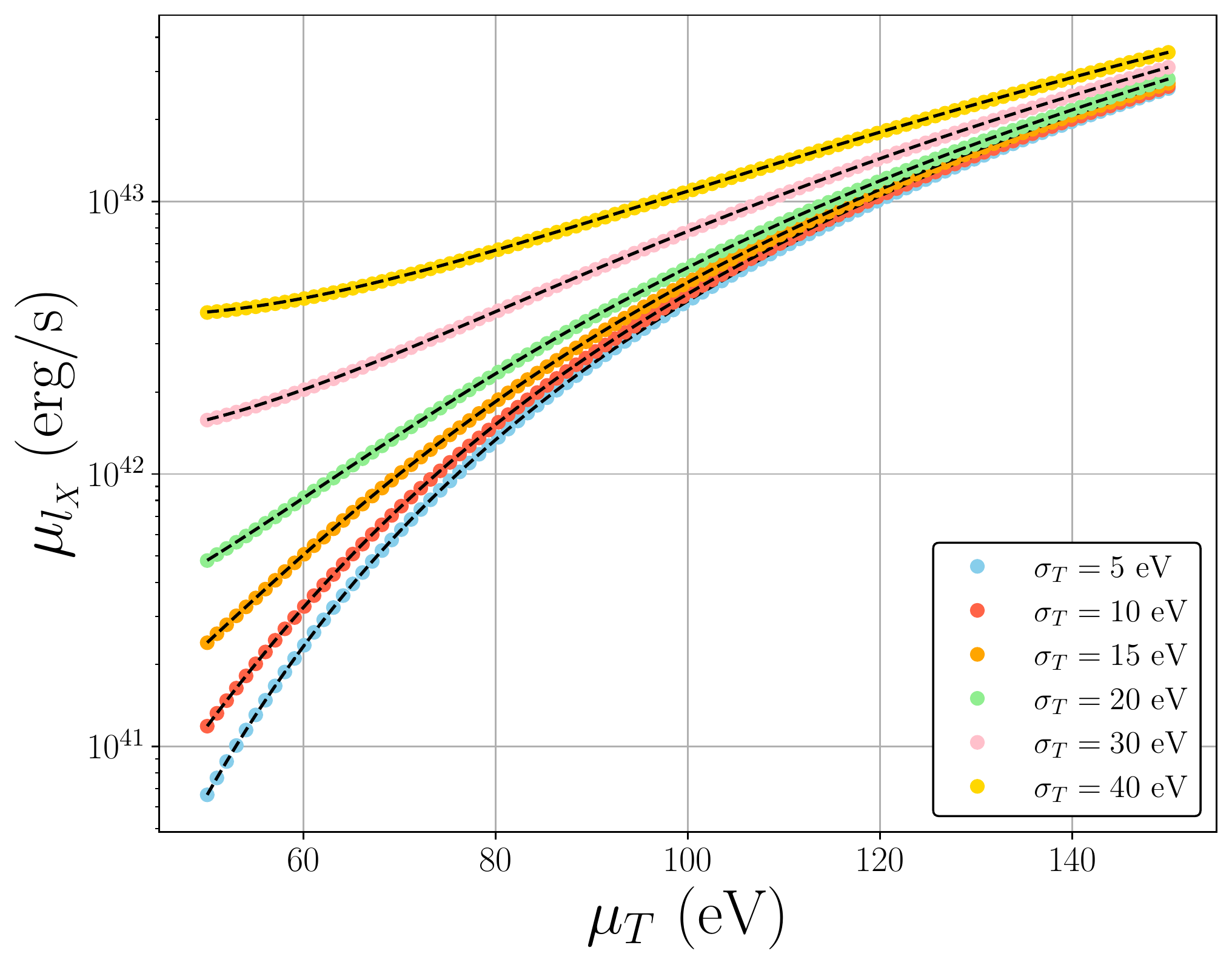}
\caption{ The mean of the X-ray luminosity distribution as a function of mean temperature parameter, for different values of $\sigma_T$. The coloured dots are numerically calculated, while the black dashed curves are  equation (\ref{xraymoms}).     }
\label{mean}
\end{figure}

\begin{figure}
\includegraphics[width=\linewidth]{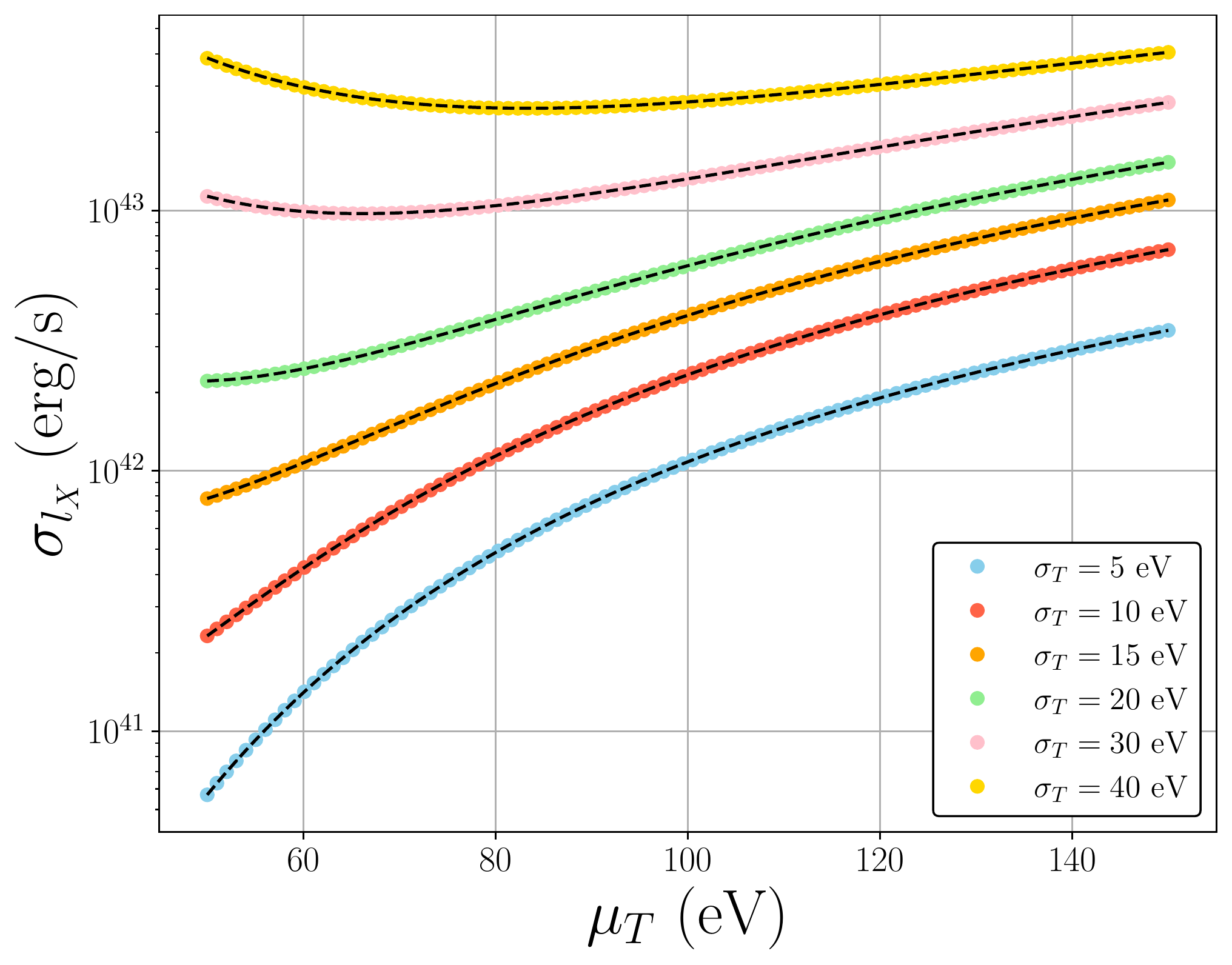}
\caption{  The standard deviation of the X-ray luminosity distribution as a function of mean temperature parameter, for different values of $\sigma_T$. The coloured dots are numerically calculated,  while the black dashed curves are equation (\ref{xraymoms}).   }
\label{sigma}
\end{figure}

The  integrand in equation (\ref{fdef})  is strongly peaked at a value of $z$  which minimises the argument of the exponent.  The value of $z$ at which this occurs, denoted $z_\star$, is given by the solution of the equation 
%\beq
%{{\rm d} \over {\rm d}z } \left[ -{(z - \mu_N - \alpha \ind \sigma_N^2)^2 \over 2\sigma_N^2}  - \alpha e^{-z} \right]_{z_\star} = 0 ,
%\eeq
%which is equivalently expressed as the following condition
\beq
z_\star = \mu_N + \alpha \ind \sigma_N^2 + \alpha \sigma_N^2 \exp(-z_\star) .
\eeq
The  solution of this equation may be expressed as:
\beq
z_\star = \mu_N + \alpha \ind \sigma_N^2 +W\left[ \alpha \sigma_N^2 \exp\left(-\mu_N - \alpha \ind \sigma_N^2\right) \right], 
\eeq
where $W$ is once again the Lambert W function defined in equation (\ref{lambdef}). The entire contribution to the integral results from a small region centred around $z_\star$, with very little contribution from other ranges of $z$ because of the strong exponential cut-off, and the function $f_\alpha$ can therefore be evaluated asymptotically using Laplace's method (Bender \& Orszag 1978).   In brief, the leading asymptotic value of the integral $I = \int\exp[- f(x)] \, {\rm d}x$ when $f(x)$ is large and has a sharp minimum, is formally {given by $I \simeq \sqrt{{2\pi / |f''(x_\star)|  }} \,\, {\exp[ -f(x_\star)]}$, where} $x_\star$ is the location of the minimum of $f(x)$, and a prime denotes a derivative with respect to $x$.   We then find:
\begin{align}\label{fapprox}
f_\alpha  &=  \int\limits_{-\infty}^{\infty} \exp\left[-{(z - \mu_N - \alpha \ind \sigma_N^2)^2 \over 2\sigma_N^2} - \alpha e^{-z} \right] {\rm d} z \nonumber \\ 
&\simeq  \sqrt{2\pi}\sigma_N \left(1 + \alpha \sigma_N^2 \exp(-z_\star)\right)^{-1/2}  \nonumber \\  
&\quad\quad \quad\quad\exp\left[-{(z_\star - \mu_N - \alpha \ind \sigma_N^2)^2 \over 2\sigma_N^2} - \alpha e^{-z_\star}\right] . %\nonumber \\ 
%&= \sqrt{2\pi}\sigma_N \left(1 + \alpha \sigma_N^2 \exp(-z_\star)\right)^{-1/2}  \phi(z_\star, \alpha) \nonumber \\  
%&\quad  \exp\left( -{1 \over 2 \sigma_N^2} \left(W\left[ \alpha \sigma_N^2 \exp\left(-\mu_N - \alpha \ind \sigma_N^2\right) \right] \right)^2\right) .
\end{align}
The moments of the  luminosity distribution are then:
%\begin{multline}\label{xraymoms}
%{I_\alpha } \approx  L_0^{\alpha} \Bigg[1 + \alpha \sigma_N^2 \exp \Big(-\mu_N -\alpha\ind\sigma_N^2 \\ - W\left[ \alpha \sigma_N^2 \exp\left(-\mu_N - \alpha \ind \sigma_N^2\right) \right]\Big)\Bigg]^{-1/2}\\
%\times \exp\Bigg[\alpha \ind \mu_N + {1\over 2} \alpha^2 \ind^2 \sigma_N^2  \\ -{1 \over 2 \sigma_N^2} \left(W\left[ \alpha \sigma_N^2 \exp\left(-\mu_N - \alpha \ind \sigma_N^2\right) \right] \right)^2 \\ - \alpha \exp\left(-\mu_N -\alpha\ind\sigma_N^2 - W\left[ \alpha \sigma_N^2 \exp\left(-\mu_N - \alpha \ind \sigma_N^2\right) \right]\right) \Bigg] .
%\end{multline}
\begin{multline}\label{xraymoms}
{I_\alpha } \simeq  L_0^{\alpha} \Bigg[1 + \alpha \sigma_N^2 \exp \Big(-\mu_N -\alpha\ind\sigma_N^2  - w_\alpha \Big)\Bigg]^{-1/2}\\
 \exp\Bigg[\alpha \ind \mu_N + {1\over 2} \alpha^2 \ind^2 \sigma_N^2   -{w_\alpha^2 \over 2 \sigma_N^2}   - \alpha \exp\left(-\mu_N -\alpha\ind\sigma_N^2 - w_\alpha \right) \Bigg] ,
\end{multline}
where we have defined 
\beq
w_\alpha \equiv W\left[ \alpha \sigma_N^2 \exp\left(-\mu_N - \alpha \ind \sigma_N^2\right) \right] .
\eeq  
An explicit numerical evaluation of the $I_\alpha$ integrals finds that equation (\ref{xraymoms}) is accurate to significantly better than $1\%$, provided that $\sigma_T \lesssim \mu_T$ and $\mu_T \lesssim T_l$ (See Figs.  [\ref{mean}], [\ref{sigma}], [\ref{rms-flux_1}--\ref{kurt_2}]). 

Figs. [\ref{mean}] and [\ref{sigma}] show the mean and standard deviation of the distribution (\ref{xray_dist}), computed numerically (coloured points) and by using equation (\ref{xraymoms}) (dashed black curves). These curves were produced for the typical tidal disruption event parameter values used for Figs. [\ref{PDF_1}--\ref{PDF_4}].  For certain regions of parameter space, the standard deviation of the luminosity distribution is larger than its mean, formally a high variance regime.

\subsection{The critical distribution-mode phase-transition curve } 
Figs. [\ref{PDF_1}] and [\ref{PDF_3}] show a remarkable property of the probability density function: at moderate to high values of $\sigma_T/\mu_T$, the luminosity probability density function has no local maximum, instead monotonically declining for all $l$. This is completely unlike the log-normal distribution, which always has finite mode at $\exp(\mu_N - \sigma_N^2)$. 

For each value of  $\mu_T$, there exists a critical value of $\sigma_T$  at which the  luminosity probability density function undergoes a dramatic transition from being centred around a particular modal value, to being a modeless distribution, with a strong tendency toward large amplitude dimming events.   We shall refer to this sharp behavioural change as a ``phase transition''.  The mode of $p_Y$ is most conveniently evaluated from $\partial_z \ln p_Y = 0$. (Recall that $z \equiv \ind^{-1} y^{-1/\ind}$.)    We have 
\begin{multline}
\ln p_Y(z) = C + \ind \ln(z) - \ln(1 + W(z)) \\ - {1 \over 2\sigma_N^2} (\ln W(z) + \ln\ind + \mu_N)^2 ,
\end{multline}
where $C$ is a constant.  Then,  $\partial_z \ln p_Y = 0$ implies
\beq
{\ind \over z} - {{\rm d} W \over {\rm d}z } \left[ {\ln\left(W \ind e^{\mu_N}\right) \over W \sigma_N^2} + {1 \over 1 + W} \right] = 0 .
\eeq
This may be simplified using equations (\ref{lambdef}) and (\ref{derivlamb}), which leads to the equation
\beq\label{modeeqn}
\ind e^{\mu_N} W = \exp\left[ \ind \sigma_N^2 \left({1 + (2 - 1/\ind) W + W^2 \over 1 + W} \right)\right]  .
\eeq
Just as $x = \exp(\beta x)$ has no real solutions for $\beta > 1/e$, equation (\ref{modeeqn}) has real solutions only for a limited range of $\sigma_T/\mu_T$.  The critical value of $\sigma_T$ for which solutions no longer exist corresponds to the point at which both the functions on the left and right side of this equation {\it and their gradients} are equal.  Denoting this point  as $W_\star$ and equating the derivatives, leaves a governing cubic for $W_\star$:
\beq
(3\ind - 1)\sigma_N^2  W_\star^3 + (2 \ind \sigma_N^2 - 1)W_\star^2 - (2+ \ind\sigma_N^2  )W_\star - 1 = 0 .
\eeq
This cubic has two roots which are close to $-1$ (this can be understood by noting that in the limit $\sigma_N \rightarrow 0$, the cubic becomes a quadratic with double root at $-1$), and one positive root, which is the root of interest. Denoting this root $W_+$, we note that the mode of the luminosity distribution can only exist in the range 
\beq
\ind^{-1} \exp(-\mu_N) < W_{\rm mode} < W_+ .
\eeq
The critical $\mu_T$-$\sigma_T$ transition curve is therefore given by
\beq
\ind e^{\mu_N}W_+ =   \exp\left[ \ind \sigma_N^2 \left({1 + (2 - 1/\ind) W_+ + W_+^2 \over 1 + W_+} \right)\right]  . 
\eeq
As $\mu_N$ and $\sigma_N$ depend only upon $\mu_T$ and $\sigma_T/\mu_T$, and $W_+$ depends only upon $\sigma_N$ (for fixed $\ind$), this condition is in reality a boundary in the $\mu_T$ - $\sigma_T/\mu_T$ parameter plane. 

\begin{figure}
\includegraphics[width=\linewidth]{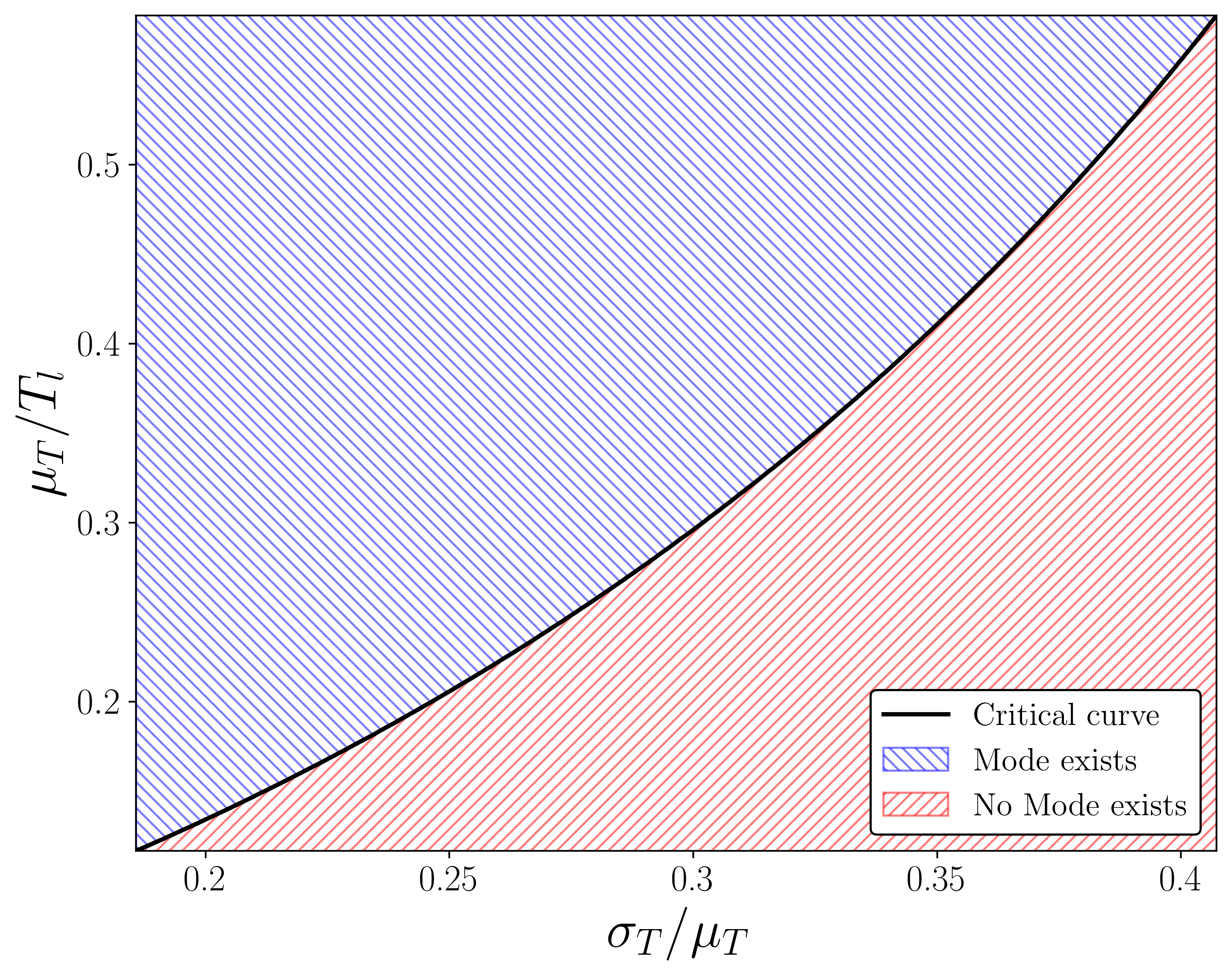}
\caption{ The critical curve in $\mu_T/T_l$--$\sigma_T/\mu_T$ parameter space. Above the black curve the luminosity distribution has a well defined mode, below the curve the distribution does not.     }
\label{CC1}
\end{figure}

\begin{figure}
\includegraphics[width=\linewidth]{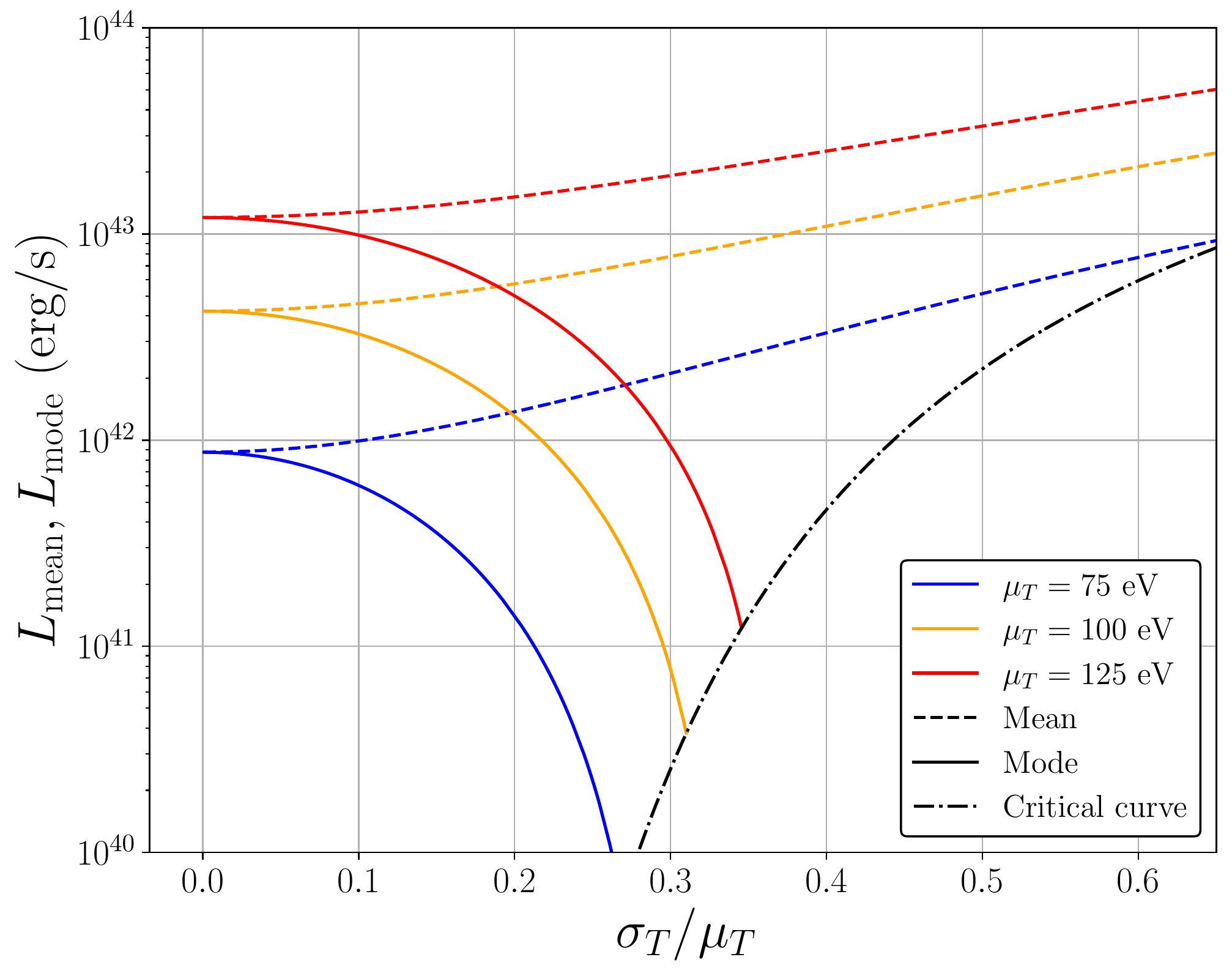}
\caption{  The mean (dashed) and mode (solid) of the luminosity probability density function, plotted against $\sigma_T/\mu_T$ for three different mean temperatures (denoted with colour). The critical transition curve is denoted by a black dot-dashed curve, beyond which the distribution has no mode.   }
\label{CC2}
\end{figure}

This critical transition curve is shown in Fig.  [\ref{CC1}]. As expected from Figs. [\ref{PDF_1}] and [\ref{PDF_3}], higher values of $\mu_T/T_l$ (hotter discs) are more stable in the sense that they undergo this transition at higher values of $\sigma_T/\mu_T$. The scale of the temperature variance required to undergo this transition is high but not at all unphysical, roughly $\sigma_T / \mu_T \sim 0.2-0.4$. 

This critical transition is plotted as a function of observed luminosity  in Fig. [\ref{CC2}]. In Fig. [\ref{CC2}] we plot the mean (dashed curves) and mode (solid curves) of the luminosity distribution, taking the typical values of parameters for comparison to tidal disruption event X-ray observations used in Figs. [\ref{PDF_1}--\ref{PDF_4}], for three different mean disc temperatures.  While the mean and mode converge in the limit $\sigma_T/\mu_T \rightarrow 0$, they diverge for increasing $\sigma_T$.  The mean always increases as a function of $\sigma_T/\mu_T$ (i.e., increasing $\sigma_N$ in equation \ref{xraymoms}), while the mode always decreases.  However, the mode only exists for a finite range of $\sigma_T/\mu_T$, and ceases to exist when the modal curve crosses a critical boundary (shown by the black dot-dashed curve in Fig. [\ref{CC2}]).

\subsection{The lack of a  linear rms-flux relation} 
The moments of the luminosity distribution are useful in interpreting observations of fluctuating disc spectra.  For instance, it is well known that many accretion disc systems, including X-ray binaries, follow a linear rms-flux relationship, i.e. the standard deviation $\sigma_{l_X}$ and mean $\mu_{l_X}$ of their X-ray light curves are found to be directly proportional to one another (e.g., Uttley {\it et al}. 2005).  

In an X-ray binary this is a direct consequence of the observed X-ray luminosity arising  from the bulk of the disc spectrum, meaning it is  well described by the log-normal distribution.   The disc systems studied in this paper, however, whose luminosity comes from the Wien tail of the spectrum,  will  {\it not}  follow the linear rms-flux relationship, as we now demonstrate.   A general prediction of this work therefore is that relevant disc systems, e.g. the cooler soft-state discs around large mass black holes observed at X-ray frequencies, will be found to deviate from the linear rms-flux relationship.  

The variance of the  luminosity distribution is defined in terms of its moments as 
\beq
\sigma_{l_X}^2 = I_2 - (I_1)^2 ,
\eeq
while the mean of the  luminosity $\mu_{l_X}$ is equal to the distributions first moment $I_1$. 
Thus, a linear rms-flux relationship requires  
\beq
{\sigma_{l_X} \over \mu_{l_X}} = {\sqrt{I_2 - (I_1) ^2 } \over I_1 } \approx {\rm cst}. 
\eeq
\begin{figure}
\includegraphics[width=\linewidth]{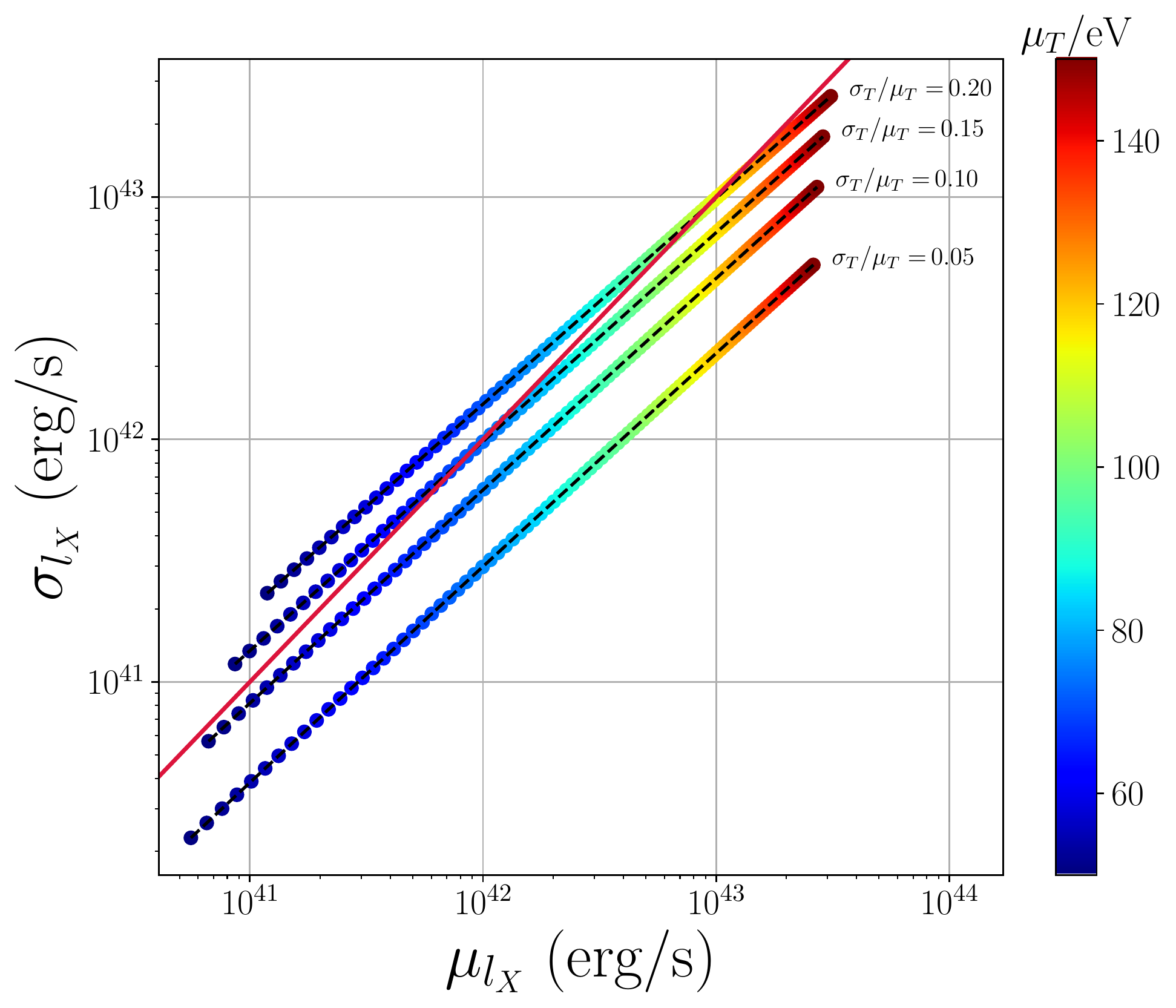}
\caption{ The standard deviation of the X-ray luminosity $\sigma_{l_X}$, plotted against its mean $\mu_{l_X}$ for different values of the temperature parameters $\mu_T$ and $\sigma_T/\mu_T$. The mean of the temperature distribution $\mu_T$ is denoted by the colour of the numerically calculated points, and the value of $\sigma_T/\mu_T$ is displayed next to each curve.  The solid crimson line displays the curve $\sigma_{l_X} = \mu_{l_X}$, and curves which are not parallel to the crimson curve do not follow the linear rms-flux relationship. }
\label{rms-flux_1}
\end{figure}

\begin{figure}
\includegraphics[width=\linewidth]{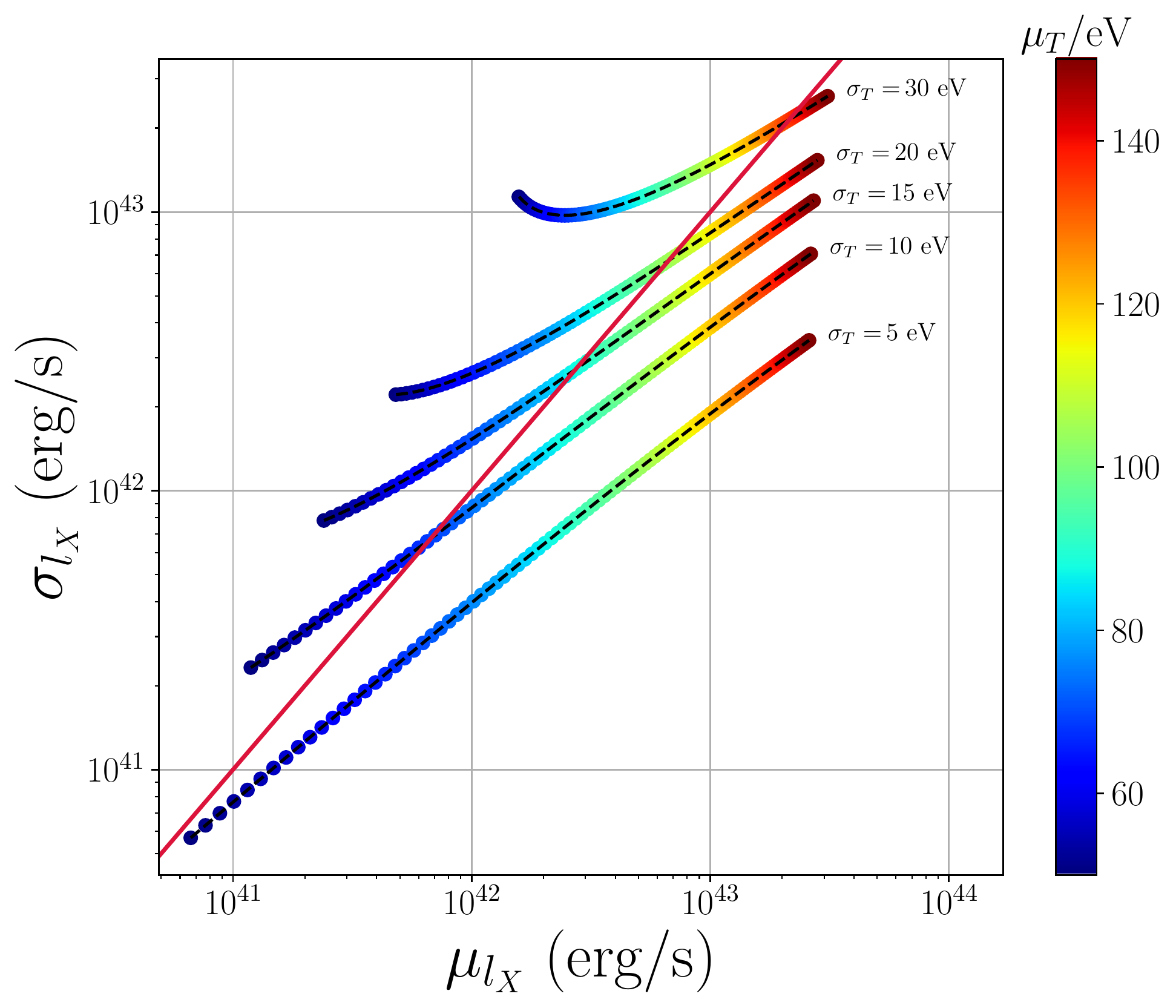}
\caption{ As in Fig. [\ref{rms-flux_1}], except for a constant temperature variance parameter $\sigma_T$, denoted on plot.  }
\label{rms-flux_2}
\end{figure}

For a lognormal distribution, the ratio $\sigma_{l_X}/\mu_{l_X}$ is only a function of $\sigma_N$. However, using the results derived above, we find for the high energy distribution the following result:
\begin{multline}\label{rmf}
\left({\sigma_{l_X} \over \mu_{l_X}}\right)^2 + 1 = {I_2 \over I_1^2 }  \approx 
A \exp\Bigg(\ind^2\sigma_N^2 - {1 \over 2 \sigma_N^2} (w_2^2 - 2w_1^2)  \\ + 2 \exp(-\mu_N)  \left[ \exp(-\ind \sigma_N^2 - w_1^2) - \exp(-2 \ind\sigma_N^2 - w_2^2) \right] \Bigg)  ,
\end{multline}
where we have defined 
\beq
A \equiv { \Bigg[1 +  \sigma_N^2 \exp \Big(-\mu_N - \ind\sigma_N^2  - w_1 \Big)\Bigg]  \over \Bigg[1 + 2 \sigma_N^2 \exp \Big(-\mu_N - 2\ind\sigma_N^2 - w_2 \Big)\Bigg]^{1/2} } .
\eeq
The key point of this analysis is that this ratio  remains a function of $\mu_N$, and in particular there remains a term proportional to $\exp(\exp(-\mu_N))$, where $\mu_N < 0$. As a  disc observed at high energies increases in brightness, the parameter $\mu_N$ must increase,  and thus, in our model, the ratio $\sigma_{l_X}/\mu_{l_X}$ will change, even if $\sigma_T/\mu_T$ is held constant.  

This can be seen explicitly in Figs. [\ref{rms-flux_1}] and [\ref{rms-flux_2}], where we plot the standard deviation of the X-ray luminosity   distribution $\sigma_{l_X}$ against its mean $\mu_{l_X}$ (again, we choose parameters here relevant for comparison to tidal disruption event discs). In Figs. [\ref{rms-flux_1}] and [\ref{rms-flux_2}] the black dashed curves are calculated using the analytical results of the previous section (equation \ref{xraymoms}).  The solid crimson line displays the curve $\sigma_{l_X} = \mu_{l_X}$.  Any points lying above the crimson line satisfy $\sigma_{l_X} > \mu_{l_X}$, a parameter space which is formally high variance.   All points are computed assuming $M_{\rm BH} = 10^6 M_\odot$. 

 In Fig. [\ref{rms-flux_1}] the standard deviation of the temperature distribution is chosen to be a fixed fraction of the temperature distributions mean (denoted on plot), while in Fig. [\ref{rms-flux_2}] the temperature variance is itself fixed.  As is clear from Figs. [\ref{rms-flux_1}] and [\ref{rms-flux_2}], the quasi-Wien X-ray luminosity of these solutions do not follow the  linear rms-flux relationship, as the curves of $\sigma_{l_X}$ versus $\mu_{l_X}$ are not parallel to the crimson $\sigma_{l_X} = \mu_{l_X}$ curve.  In addition to not following the linear rms-flux relationship, Figs. [\ref{rms-flux_1}] and [\ref{rms-flux_2}] demonstrate that the X-ray luminosity of these discs will, for a large region of parameter space, be formally high variance. By this we mean that the coefficient of variance, defined as $c_V \equiv \sigma_{l_X} / \mu_{l_X}$,  is greater than unity.  In this region large amplitude outliers are expected to be common.   Perhaps the simplest test of the theory developed in this paper therefore would be to compute the evolving standard deviation and mean of TDE X-ray light curves, and test for deviations from a linear relationship.

\subsection{Large amplitude outliers: the kurtosis of the  luminosity distribution }
As is clear from Figs. [\ref{mean}--\ref{rms-flux_2}], the probability distribution derived in this paper is notable for its extended tails.   The likelihood of observing a large amplitude outlier from a given probability distribution is normally quantified by calculating the kurtosis of that distribution.   
The kurtosis of the distribution of a quantity $X$ (with corresponding probability density $p_X$) is defined as
\beq
\kappa_X \equiv \left\langle \left({X - \mu \over \sigma}\right)^4 \right\rangle = \int\limits_{-\infty}^{\infty} p_X(x)  \left({x - \mu \over \sigma}\right)^4 \, {\rm d}x ,
\eeq
where $\mu$ and $\sigma$ are the mean and standard deviation of the variable $X$ respectively. The sense in which the kurtosis measures the ``tailedness'' of a distribution  (the propensity of a distribution to produce outliers) can be understood intuitively  by noting that all points that lie within one standard deviation of the mean ($\left| X - \mu \right| < \sigma$) contribute very little to the kurtosis, while any points which lie far from the mean $\left| X - \mu \right| \gg \sigma$ contribute a large amount.  The normal distribution has a kurtosis of 3, and thus the ``excess kurtosis'', defined as 
\beq
\hat \kappa \equiv \kappa - 3 ,
\eeq
is a more commonly used statistic. Any positive value of the excess kurtosis $\hat \kappa > 0$ is associated with a propensity to produce outliers, while $\hat \kappa \gg 1$ indicates that a probability distribution will result in many values being observed very far from the mean.  

\begin{figure}
\includegraphics[width=\linewidth]{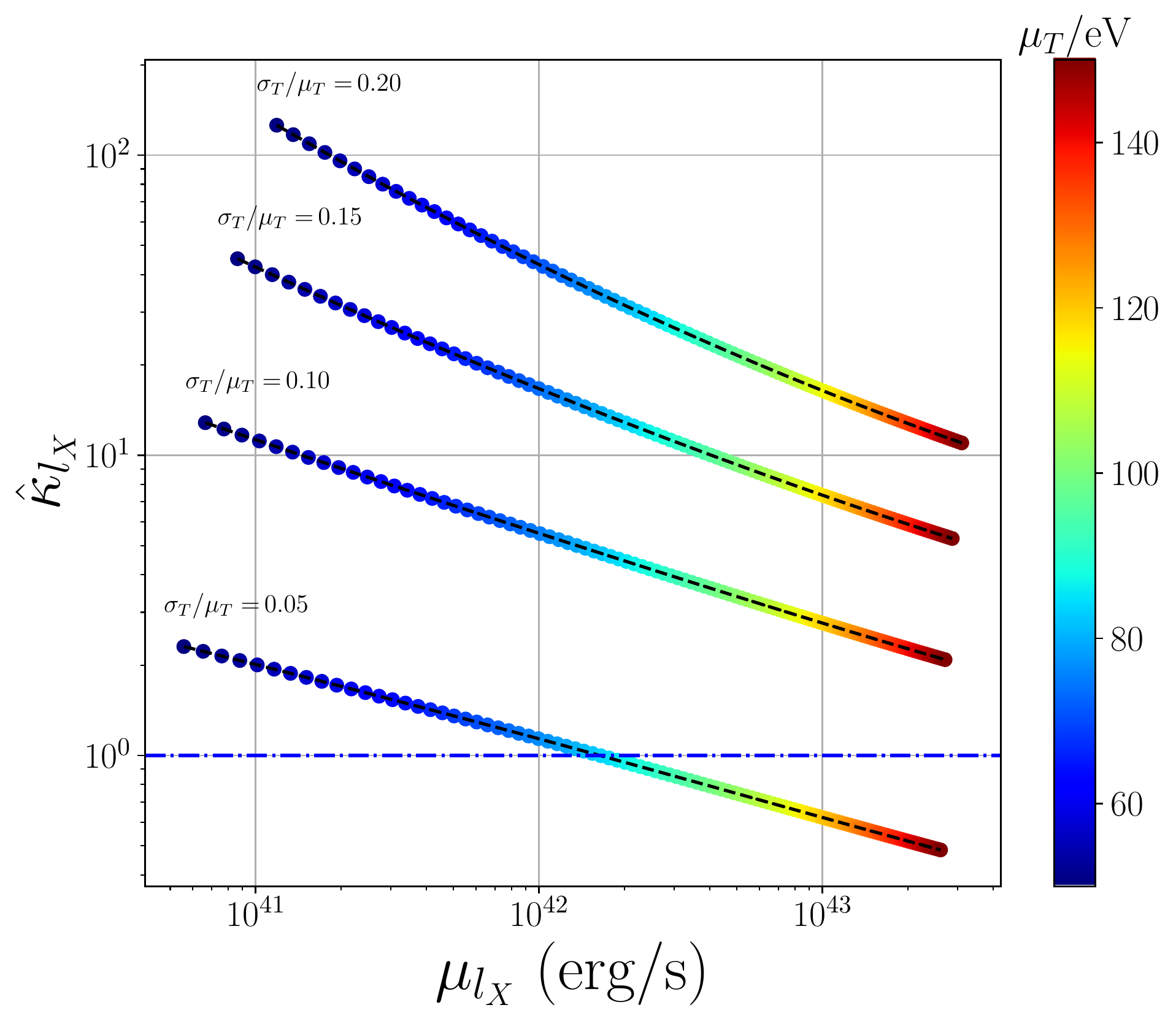}
\caption{ The excess kurtosis of the X-ray luminosity $\hat \kappa_{l_X}$, plotted against its mean $\mu_{l_X}$ for different values of the temperature parameters $\mu_T$ and $\sigma_T/\mu_T$. The mean of the temperature distribution $\mu_T$ is denoted by the colour of the numerically calculated points, and the value of $\sigma_T/\mu_T$ is displayed next to each curve.     }
\label{kurt_1}
\end{figure}

\begin{figure}
\includegraphics[width=\linewidth]{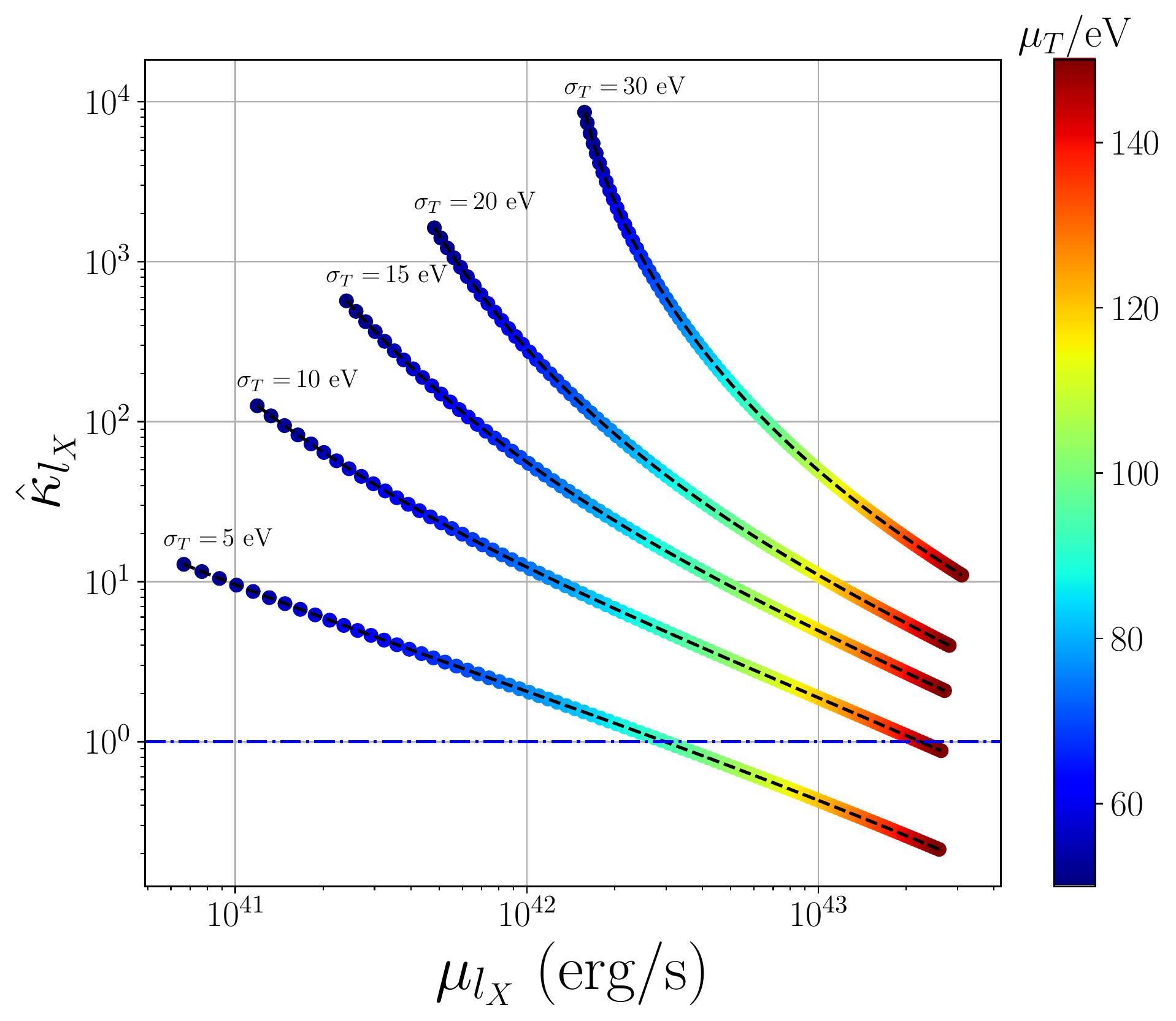}
\caption{ The excess kurtosis of the X-ray luminosity $\hat \kappa_{l_X}$, plotted against its mean $\mu_{l_X}$ for different values of the temperature parameters $\mu_T$ and $\sigma_T$. The mean of the temperature distribution $\mu_T$ is denoted by the colour of the numerically calculated points, and the value of $\sigma_T$ is displayed next to each curve.   }
\label{kurt_2}
\end{figure}

In terms of the moments of the distribution, the kurtosis of the high energy luminosity may be written in terms of the $I_\alpha$ quantities
\beq
\kappa_{l} =  \left\langle \left({L_{\rm band} - \mu_{l} \over \sigma_{l}}\right)^4 \right\rangle = {I_4 - 4 I_3 I_1 + 6 I_2 I_1 ^2 - 3 I_1^4 \over (I_2 - I_1^2)^2 } ,
\eeq
 In Figs. [\ref{kurt_1}] and [\ref{kurt_2}] the excess kurtosis is plotted as a function of the mean of the  luminosity distribution, for parameter values typical of tidal disruption event systems, for different values of the temperature parameters $\sigma_T$ and $\mu_T$. 

In Fig. [\ref{kurt_1}], the excess kurtosis is calculated for different fixed values of the ratio $\sigma_T/\mu_T$, denoted on plot. The coloured points are computed numerically, while the black dashed curves are the analytical results of equation (\ref{xraymoms}). The first point to note is that all values of the excess kurtosis are positive, indicating that outliers in this luminosity distributions  will be common. More importantly however,  all points lying above the blue dot-dashed curve have $\hat \kappa_{l_X} > 1$ and represent regions of parameter space characterised by a high propensity for large amplitude outliers.  Indeed,  the excess kurtosis can, for reasonable values of parameter space, be extremely large, of order $\gtrsim 10^2$. An excess kurtosis this high means that the tails of the  luminosity distribution are  ``super-Gaussian'': they will produce large amplitude outliers at far higher frequencies than might naively be expected,  and standard statistical fitting procedures which rely on assumptions about the normality of the errors on data points will lead to incorrect inferences. If, rather than $\sigma_T / \mu_T$, the temperature variance parameter $\sigma_T$ is fixed, then the excess kurtosis can reach even larger values, as demonstrated in Fig. [\ref{kurt_2}].

The typical values of the excess kurtosis of the high energy luminosity distribution are extremely large, and therefore it is a natural expectation that those disc systems observed in the Wien tail will frequently exhibit large amplitude outliers.  Equivalently, it is important not to over interpret observed fluctuations in e.g., TDE X-ray light curves. A fluctuation of an order of magnitude in the observed X-ray luminosity by no means corresponds to an order of magnitude fluctuation in the intrinsic physical disc structure. 

\section{Variability of the high energy disc spectrum}
The results of the previous section concern the statistical properties of the fluctuations in the integrated  luminosity across an instrument's bandpass.  The properties of the variability of a disc spectrum, as observed at a particular frequency, can also be analysed in an identical manner.  The observed spectrum of an accretion disc, as observed at a high energy  $\nu \gg kT_p/h$, is given to leading order by (Mummery \& Balbus 2021a)
\begin{equation}\label{MB2}
\nu L_{\nu}  = L_0 \left({\nu \over \nu_l}\right)^4 \left(\frac{k \T_p}{h \nu} \right)^{\ind-1} \exp\left(- \frac{h\nu}{k \T_p} \right) ,
\end{equation} 
where all of the variables have the same definitions as previously.  Thus, the dimensionless disc spectrum is defined by a function of the form 
\beq
Y = X^{\ind - 1} \exp\left(-{1/ X}\right) ,
\eeq
where  $X \equiv k\T_p / h\nu$. The probability distribution of the disc spectrum is therefore once again given by equation (\ref{xray_dist}), but with the substitutions
\begin{align}
\ind &\rightarrow \ind - 1, \\
L_0 &\rightarrow L_0 (\nu/\nu_l)^4 , 
\end{align}
and it is important to remember that $\mu_N$ is a function of observing frequency 
\beq
\mu_N = \ln\left({k \mu_T \over h \nu} {\mu_T \over \sqrt{\mu_T^2 + \sigma_T^2}}\right) .
\eeq
The moments of the probability distribution of the high energy disc spectrum are similarly given by the results of the previous section, with the above substitutions. 

\begin{figure}
\includegraphics[width=\linewidth]{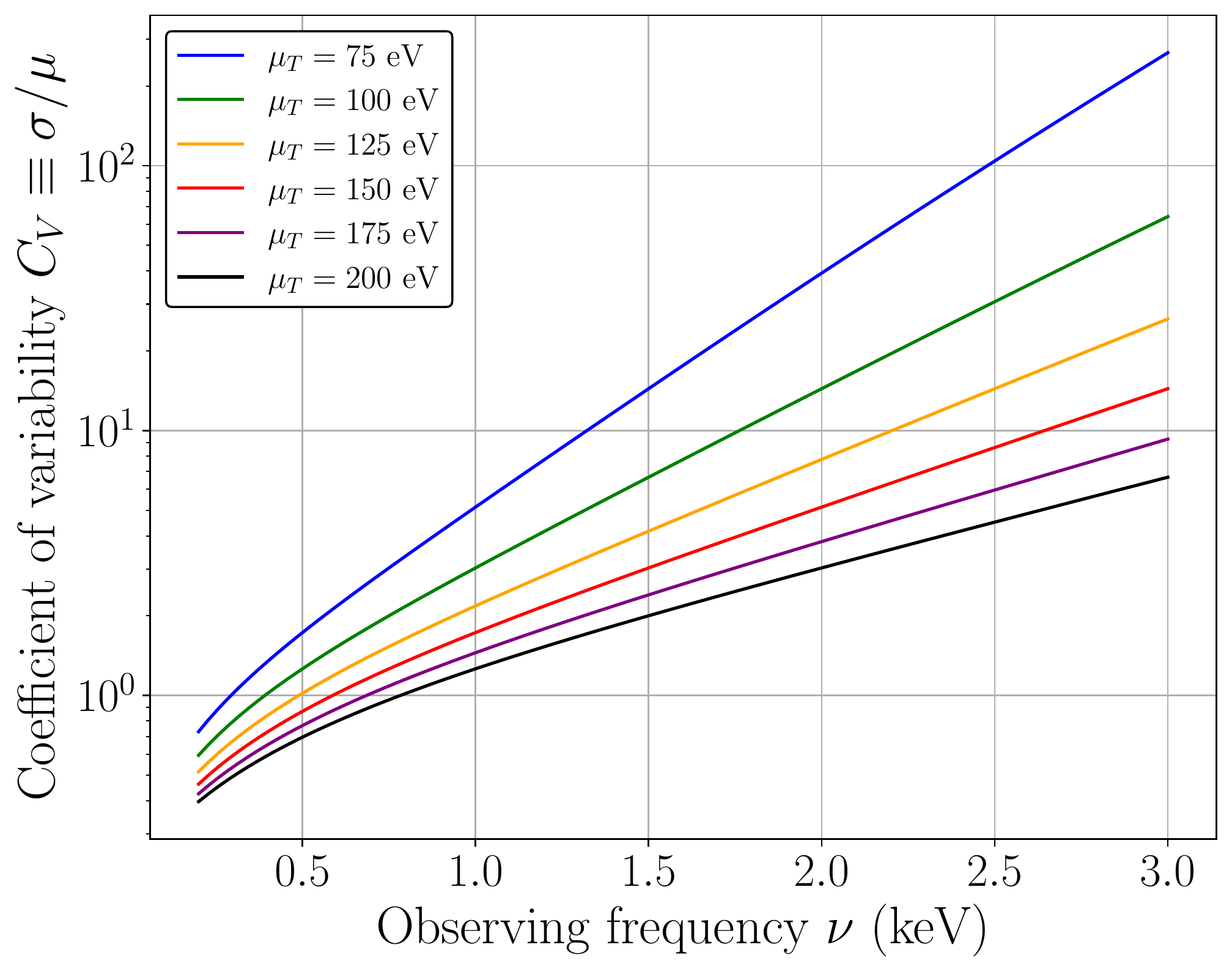}
\caption{ The coefficient of variability, defined as the ratio of the standard deviation to mean, of the high energy disc spectrum as a function of observing frequency. The mean disc temperature is displayed in the legend, and the standard deviation of the disc temperature is {$\sigma_T / \mu_T =0.2$. }   }
\label{var_freq}
\end{figure}

A key  property of this analysis is highlighted in equation (\ref{rmf}),  which shows that the ratio of the standard deviation and mean of the probability distribution  has a function dependence on $\mu_N$.  At fixed disc temperature $\mu_T$,  increasing the observing frequency $\nu$ causes $\mu_N$ to become more negative. As the fractional variability $\sigma/\mu$ has a dependence on $\exp(\exp(-\mu_N))$, this means that the fractional variability of the disc spectrum will, according to equation (\ref{rmf}),  {\it increase with increasing observing frequency}. This result, which is verified in Fig. [\ref{var_freq}],  is distinct from the properties of the log-normal distribution, and therefore represents an additional test of the theory developed in this paper.   

\section{Conclusions}
In this paper we have derived the  closed form probability density function of the accretion disc luminosity in the Wien portion of its spectrum. This was derived assuming  a log-normally distributed temperature profile, an assumption with both observational (Uttley {\it et al}. 2005) and theoretical (Turner \& Reynolds 2021) backing.  This probability density function has a number of interesting properties not shared by  the log-normal distribution, which can be examined using both asymptotic expressions for its moments (equation [\ref{xraymoms}]), and an analysis of the modal behaviour of the distribution (Fig. [\ref{CC2}]). 

An important result of this analysis is that the Wien-tail luminosity does not follow the linear rms-flux relationship associated with the bolometric luminosity of the disc.   Mathematically this results from the additional exponential amplification of the temperature fluctuations by the Wien tail of the disc spectrum, which modifies the probability distribution from log-normal.  The deviations from the linear rms-flux relationship can be relatively slight (Fig. [\ref{rms-flux_1}]), or significant (Fig. [\ref{rms-flux_2}]) depending on the exact properties of the temperature fluctuations.  Observations of the rms-flux relationship of tidal disruption events observed at X-ray energies, which satisfy the governing $h \nu_l \gg k\T_p$ assumption of the model, will represent an interesting test of the theory developed in this paper. 

A further result of  significance for  observations  is highlighted in Figs.  [\ref{rms-flux_1}], [\ref{rms-flux_2}], [\ref{kurt_1}] and [\ref{kurt_2}]: the high  energy emission from a disc is likely to be highly variable, with many large amplitude outliers observed far from the mean. In Figs. [\ref{rms-flux_1}] and [\ref{rms-flux_2}] we show that, for  reasonable regions of parameter space, the coefficients of variation $c_V \equiv \sigma_{l_X} / \mu_{l_X}$ of the luminosity distribution is in excess of unity, while Figs. [\ref{kurt_1}] and [\ref{kurt_2}] show that the high photon energy luminosity distribution can have kurtosis in excess of $\sim 100$. The tails of the distribution are extremely extended.   This fractional variability {\it increases with increasing observing frequency}, a prediction which can be  used as a further test of this analysis.  

In fact, this probability distribution  undergoes a sort of ``phase transition'' from a distribution that is centred about a modal value to one that is mode-less, at some critical value of the underlying temperature variance. Over the course of this transition, the mean of the distribution monotonically increases (Fig. [\ref{CC2}]).  A distribution with large mean and a global maxima at $0$ will clearly {exhibit an} extremely high variance,  with a notable propensity for producing large-amplitude {\it dimming} events (observational levels far below the mean). 

The mathematical results in this paper will be useful as an aid to understanding the variability properties of disc systems observed at energies higher than their peak disc temperature.  A particularly interesting example of one such system is that of tidal disruption events (with typical peak temperature scales $kT_p \sim 50-150$ eV) observed at X-ray energies ($h \nu > 300$ eV). In a follow-up analysis we shall examine in detail the X-ray variability of a number of well observed tidal disruption events, demonstrating that while the log-normal distribution provides a rather poor description of their luminosity variability, the distribution derived here is significantly more accurate.

\label{lastpage}

\section*{Acknowledgments} 
It is a pleasure to acknowledge stimulating discussions with R. Blandford, R. Fender, A. Mushtukov, and A. Ingram. We are particularly grateful to our referee C. Reynolds, for his support and valuable advice.  
This work is partially supported by the Hintze Family Charitable Trust and STFC grant ST/S000488/1.

\section*{Data accessibility statement}
No observational  data was used in producing this manuscript.

\end{document}